\documentclass[prd,superscriptaddress,amsfonts,amssymb,amsmath,showpacs,twocolumn]{revtex4-2}
\usepackage{bm}
\usepackage{amsfonts}
\usepackage{latexsym}
\usepackage[latin1]{inputenc}
\usepackage{graphicx}
\usepackage{amsmath}
\usepackage{palatino}
\usepackage{mathpazo}
\usepackage{textcomp}
\usepackage{float}
\usepackage{booktabs}
\usepackage{dcolumn}
\usepackage{hyperref}
\hypersetup{colorlinks,citecolor=blue}
\usepackage{amsmath}
\usepackage{xcolor}
\usepackage{orcidlink}
\usepackage[caption=false]{subfig}
\usepackage{commath}
\def\jnl@style{\it}
\def\aaref@jnl#1{{\jnl@style#1}}

\def\aaref@jnl#1{{\jnl@style#1}}

\def\aj{\aaref@jnl{AJ}}                   
\def\apj{\aaref@jnl{ApJ}}                 
\def\apjl{\aaref@jnl{ApJ}}                
\def\apjs{\aaref@jnl{ApJS}}               
\def\apss{\aaref@jnl{Ap\&SS}}             
\def\aap{\aaref@jnl{A\&A}}                
\def\aapr{\aaref@jnl{A\&A~Rev.}}          
\def\aaps{\aaref@jnl{A\&AS}}              
\def\mnras{\aaref@jnl{Mon.~Not.~Roy.~Astron.~Soc.}}             
\def\prd{\aaref@jnl{Phys.~Rev.~D}}        
\def\prc{\aaref@jnl{Phys.~Rev.~C}}  
\def\prl{\aaref@jnl{Phys.~Rev.~Lett.}}    
\def\qjras{\aaref@jnl{QJRAS}}             
\def\skytel{\aaref@jnl{S\&T}}             
\def\ssr{\aaref@jnl{Space~Sci.~Rev.}}     
\def\zap{\aaref@jnl{ZAp}}                 
\def\nat{\aaref@jnl{Nature}}              
\def\aplett{\aaref@jnl{Astrophys.~Lett.}} 
\def\apspr{\aaref@jnl{Astrophys.~Space~Phys.~Res.}} 
\def\physrep{\aaref@jnl{Phys.~Rep.}}      
\def\physscr{\aaref@jnl{Phys.~Scr}}       
\def\commat{\aaref@jnl{Comm.~Math.~Phys.}}              
\def\science{\aaref@jnl{Science}}               
\def\cqg{\aaref@jnl{Classical Quant.~Grav.}}            
\def\jpcs{\aaref@jnl{JPCS}}                                     
\def\ijmpd{\aaref@jnl{Int.~J.~Mod.~Phys.~D}}                    
\def\grg{\aaref@jnl{Gen.~Relat.~Gravit.}}               
\def\rpp{\aaref@jnl{Rep.~Prog.~Phys.}}          
\def\npa{\aaref@jnl{Nucl.~Phys.~A}}        
\def\lrr{\aaref@jnl{Living Rev.~Rel.}}                   
\def\jcap{\aaref@jnl{J.~Cosmology Astropart.~Phys.}}    
\def\rmp{\aaref@jnl{Rev.~Mod.~Phys.}}   
\def\epjc{\aaref@jnl{Eur.~Phys.~J.~C}}

\allowdisplaybreaks[1]

\begin{document}

\color{black}       

\title{Constraining the cosmological model of modified $f(Q)$ gravity: Phantom dark energy and observational insights}

\author{M. Koussour\orcidlink{0000-0002-4188-0572}}
\email[Email:]{pr.mouhssine@gmail.com}
\affiliation{Quantum Physics and Magnetism Team, LPMC, Faculty of Science Ben
M'sik,\\
Casablanca Hassan II University,
Morocco.}

\author{N. Myrzakulov\orcidlink{0000-0001-8691-9939}}
\email[Email: ]{nmyrzakulov@gmail.com}
\affiliation{L. N. Gumilyov Eurasian National University, Astana 010008,
Kazakhstan.}
\affiliation{Ratbay Myrzakulov Eurasian International Centre for Theoretical
Physics, Astana 010009, Kazakhstan.} 

\author{Alnadhief H. A. Alfedeel\orcidlink{0000-0002-8036-268X}}%
\email[Email:]{aaalnadhief@imamu.edu.sa}
\affiliation{Department of Mathematics and Statistics, Imam Mohammad Ibn Saud Islamic University (IMSIU),\\
Riyadh 13318, Saudi Arabia.}
\affiliation{Department of Physics, Faculty of Science, University of Khartoum, P.O. Box 321, Khartoum 11115, Sudan.}
\affiliation{Centre for Space Research, North-West University, Potchefstroom 2520, South Africa.}

\author{Amare Abebe}
\email[Email: ]{amare.abebe@nithecs.ac.za}
\affiliation{Centre for Space Research, North-West University, Potchefstroom 2520, South Africa.}
\affiliation{National Institute for Theoretical and Computational Sciences (NITheCS), 3201 Stellenbosch, South Africa}

%
\date{\today}
\begin{abstract}
Despite the significant accomplishments of general relativity, numerous unresolved issues persist in our understanding of the cosmos. One of the most perplexing challenges is the ongoing accelerated expansion of the Universe, which continues to elude a complete explanation. Consequently, scientists have proposed various alternative theories to GR in pursuit of a deeper understanding. In our analysis, we delve into the recently proposed modified $f(Q)$ gravity, where $Q$ represents the non-metricity scalar responsible for gravitational effects. Specifically, we investigate a cosmological model characterized by the functional form $f(Q) = Q+\alpha Q^n$, where $\alpha$ (with $\alpha \neq 0$) and $n$ serve as free parameters. Utilizing this functional form, we construct our Hubble rate, incorporating a specific equation of state to describe the cosmic fluid. Furthermore, we leverage a dataset consisting of 31 data points from Hubble measurements and an additional 1048 data points from the Pantheon dataset. These data serve as crucial constraints for our model parameters, and we employ the Markov Chain Monte Carlo (MCMC) method to explore the parameter space and derive meaningful results. With our parameter values constrained, our analysis yields several noteworthy findings. The deceleration parameter suggests a recent accelerated phase in the cosmic expansion. In addition, the EoS parameter paints a portrait of dark energy exhibiting phantom-like characteristics. Furthermore, we delve into the application of cosmological diagnostic tools, specifically the statefinder and the $Om(z)$ diagnostics. Both of these tools align with our previous conclusions, confirming the phantom-like behavior exhibited by our cosmological model. These results collectively contribute to our understanding of the dynamic interplay between gravity, dark energy, and the expanding cosmos.

\end{abstract}

\maketitle

\date{\today}
\section{Introduction}

\label{sec1}

Recent observations in modern cosmology, including SN Ia (type Ia Supernova) \cite{SN1, SN2}, LSS (Large Scale Structure) \cite{LS1, LS2}, the WMAP (Wilkinson Microwave Anisotropy Probe) experiment \cite{WMAP1, WMAP2, WMAP9}, CMB (Cosmic Microwave Background) \cite{CMB1, CMB2}, and BAOs (Baryonic Acoustic Oscillations) \cite{BAO1, BAO2}, have conclusively shown that our Universe is undergoing accelerated expansion. Furthermore, these cosmological observations indicate that the visible matter we observe constitutes only 5\% of the total content of the Universe, with the remaining 95\% existing in the form of unknown components referred to as Dark Energy (DE) and dark matter. However, these findings are at odds with General Relativity (GR), particularly the well-known Friedmann equations, which are derived from the application of GR to a homogeneous and isotropic Universe described by the Friedmann-Lema\^itre-Robertson-Walker (FLRW) space-time. Consequently, it is evident that GR cannot serve as the ultimate theory of gravity and may instead represent a special case within a more comprehensive theory.

To address the observations regarding the accelerated expansion of the Universe, several alternatives have been proposed. One such alternative within the framework of GR is the inclusion of a new energy component, known as DE, characterized by a large negative pressure. The cosmological constant ($\Lambda$) introduced by Einstein in his field equations is currently the leading candidate for DE, as it aligns well with observations. The prevailing idea is that the source of $\Lambda$ is vacuum energy predicted by quantum theory \cite{Weinberg}. However, this idea faces two main challenges. The first is the fine-tuning problem, which arises due to the substantial difference between theoretical and experimental values. The second challenge is the coincidence problem, which questions why the energy density of $\Lambda$ remains constant despite cosmological observations indicating that the sources of DE vary slowly with cosmic time. The latter issue can be resolved by introducing a time-variable cosmological constant through the inclusion of a scalar field with kinetic and potential terms, as seen in the quintessence DE model \cite{quintessence}. Other dynamical models of DE, such as phantom DE \cite{phantom}, k-essence \cite{kessence}, chameleon \cite{Chameleon}, tachyon \cite{tachyon}, Chaplygin gas \cite{Cgas1, Cgas2}, and little sibling of the big rip \cite{bouhmadi2017,bouhmadi2018}, have also been proposed.

The second alternative involves modifying Einstein's theory of GR. In GR, curvature is described by the Ricci scalar $R$, based on Riemannian geometry. Modified $f(R)$ gravity replaces the Ricci scalar with general functions of $R$ \cite{fR}. Additionally, there are other alternatives to GR, such as $f(T)$ gravity, where gravitational effects are described by the concept of torsion $T$ \cite{fT}. Recently, a new gravity theory based on Weyl geometry, which is more general than Riemannian geometry, has been proposed. This theory, known as $f(Q)$ gravity, describes gravitational effects in terms of non-metricity, which represents the variation of vector length during parallel transport \cite{Jimenez1, Jimenez2}. In Weyl geometry, the covariant derivative of the metric tensor is not zero but mathematically determined by the non-metricity tensor, denoted as $Q_{\gamma \mu
\nu }=-\nabla _{\gamma }g_{\mu \nu }$ \cite{Xu}. Energy conditions and cosmography in $f(Q)$ gravity have been explored by Mandal et al. \cite{Mandal1, Mandal2}, while Harko et al. investigated matter coupling in modified $Q$ gravity assuming a power-law function \cite{Harko1}. Dimakis et al. discussed quantum cosmology for a polynomial $f(Q)$ model \cite{Dimakis}, and other related works include \cite{Shekh, Koussour1, Koussour2}.

In the literature, the Equation of State (EoS) parameter is commonly employed to characterize the nature of DE in various models. The EoS parameter represents the relationship between the pressure and the energy density of the Universe. Its value varies depending on the specific model under consideration. For instance, in a matter-dominated Universe, the EoS parameter is typically $\omega = 0$, while for a radiation-dominated Universe, it takes the value $\omega = 1/3$. In the case of an accelerating expanding Universe, the EoS parameter $\omega$ assumes different values: $\omega = -1$ corresponds to a cosmological constant, $-1 < \omega < -1/3$ corresponds to quintessence (a type of DE), and $\omega < -1$ corresponds to phantom DE \cite{Koussour3, Koussour4}. The current value of the EoS parameter, as reported by the Planck Collaboration, is $\omega _{0}=-1.028\pm
0.032$ \cite{Planck2015, Planck2018}. In this study, we adopt a model-independent approach \cite{Pacif, Chevallier, Campo} and incorporate an effective EoS parameter to account for the current acceleration of the Universe within the framework of $f(Q)$ gravity. To constrain the model parameters, we use two recent sets of observational data: Hubble $Hz$ measurements and the $Pantheon$ datasets. The Hubble datasets consist of 31 data points obtained through the differential age method \cite{Yu/2018,Moresco/2015}. In addition, the $Pantheon$ datasets, spanning the redshift range $0.01<z<2.3$, provide 1048 data points \cite{Scolnic}. We employ the Markov Chain Monte Carlo (MCMC) method \cite{Mackey} to estimate the model parameters. Moreover, we employ two diagnostic tools to discern between different DE models. Firstly, we consider the statefinder parameters $\left( r,s\right)$ introduced by Sahni et al. \cite{Sahni, Alam}. For example, in the case of the statefinder parameters, the $\Lambda$CDM model corresponds to $\left( r=1, s=0 \right)$, the Holographic DE model corresponds to $\left( r=1, s=\frac{2}{3} \right)$, the Chaplygin gas model corresponds to $\left( r>1, s<0 \right)$, and the quintessence model corresponds to $\left( r<1, s>0 \right)$. Secondly, we employ the $Om(z)$ diagnostic introduced in \cite{Sahni1}. The $Om(z)$ diagnostic relies on the slope of the function $Om(z)$, where a negative slope indicates quintessence behavior, a positive slope indicates phantom behavior, and a zero slope corresponds to $\Lambda$CDM.

The paper is structured as follows. Sec. \ref{sec2} provides a brief overview of the mathematical formalism of $f(Q)$ gravity in a flat FLRW Universe. In Sec. \ref{sec3}, we present a specific $f(Q)$ cosmological model and derive the Hubble parameter by incorporating an effective EoS parameter. The observational constraints on the model parameters are discussed in Sec. \ref{sec4}, using the $Hz$ datasets consisting of 31 data points and the $Pantheon$ datasets consisting of 1048 data points. Additionally, the behavior of cosmological parameters, including the deceleration parameter and EoS parameter, is analyzed in this section. Secs. \ref{sec5} and \ref{sec6} are dedicated to the examination of geometrical parameters. Sec. \ref{sec5} focuses on the statefinder parameters, while Sec. \ref{sec6} introduces the $Om(z)$ diagnostic tool. Finally, Sec. \ref{sec7} summarizes the conclusions drawn from the study.

\section{$f(Q)$ gravity theory}
\label{sec2}

In the realm of differential geometry, the metric tensor $g_{\mu \nu }$ is regarded as a generalization of gravitational potentials. Its primary function is to determine angles, distances, and volumes. On the other hand, the affine connection $\Upsilon {^{\gamma }}_{\mu \nu }$ plays a crucial role in parallel transport and covariant derivatives. In the context of Weyl geometry, which incorporates the non-metricity term $Q$, the Weyl connection $\Upsilon {^{\gamma }}_{\mu \nu }$ can be decomposed into two distinct components: the Christoffel symbol ${\Gamma
^{\gamma }}_{\mu \nu }$ and the disformation tensor ${L^{\gamma }}_{\mu \nu }$. This decomposition allows for a better understanding of the geometric properties and interactions within Weyl geometry \cite{Xu},
\begin{equation}
\Upsilon {^{\gamma }}_{\mu \nu }={\Gamma ^{\gamma }}_{\mu \nu }+{L^{\gamma }}%
_{\mu \nu },  \label{WC}
\end{equation}%
where the Christoffel symbol is determined in terms of the metric tensor $g_{\mu \nu }$ by
\begin{equation}
{\Gamma ^{\gamma }}_{\mu \nu }\equiv \frac{1}{2}g^{\gamma \sigma }\left(
\partial _{\mu }g_{\sigma \nu }+\partial _{\nu }g_{\sigma \mu }-\partial
_{\sigma }g_{\mu \nu }\right) 
\end{equation}%
and the disformation tensor ${L^{\gamma }}_{\mu \nu }$ is obtained from the non-metricity tensor $Q_{\gamma \mu \nu }$ as
\begin{equation}
{L^{\gamma }}_{\mu \nu }\equiv \frac{1}{2}g^{\gamma \sigma }\left( Q_{\nu
\mu \sigma }+Q_{\mu \nu \sigma }-Q_{\gamma \mu \nu }\right) .  \label{L}
\end{equation}%

The non-metricity tensor $Q_{\gamma \mu \nu }$ is defined as the covariant derivative of the metric tensor with respect to the Weyl connection $\Upsilon 
{^{\gamma }}_{\mu \nu }$, expressed as 
\begin{equation}
Q_{\gamma \mu \nu }=-\nabla _{\gamma }g_{\mu \nu },
\end{equation}%
and it can be calculated by 
\begin{equation}
Q_{\gamma \mu \nu }=-\partial _{\gamma }g_{\mu \nu }+g_{\nu \sigma }\Upsilon 
{^{\sigma }}_{\mu \gamma }+g_{\sigma \mu }\Upsilon {^{\sigma }}_{\nu \gamma
}.
\end{equation}%

The theoretical framework employed in this study is symmetric teleparallel gravity, also known as $f(Q)$ gravity, which is equivalent to the well-known theory of gravity (GR) \cite{Jimenez1}. The equivalence between $f(Q)$ gravity and GR is established in the coincident gauge, where the Weyl connection is set to zero, $\Upsilon {^{\gamma }}_{\mu \nu }=0$. In this gauge, the curvature tensor also becomes zero, resulting in a flat spacetime geometry. Consequently, the covariant derivative $%
\nabla _{\gamma }$ simplifies to the partial derivative $\partial _{\gamma }$, leading to the expression $Q_{\gamma \mu \nu }=-\partial _{\gamma }g_{\mu \nu }$.

From the preceding discussion, the Levi-Civita connection ${\Gamma ^{\gamma }}%
_{\mu \nu }$ can be expressed in terms of the disformation tensor ${L^{\gamma }%
}_{\mu \nu }$ as ${\Gamma ^{\gamma }}_{\mu \nu }=-{L^{\gamma }}_{\mu \nu }$.

The action for symmetric teleparallel gravity is defined as \cite%
{Jimenez1, Jimenez2} 
\begin{equation}
S=\int \sqrt{-g}d^{4}x\left[ -\frac{1}{2}f(Q)+\mathcal{L}_{m}\right] ,
\label{action}
\end{equation}%
where $f(Q)$ is an arbitrary function of the non-metricity scalar $Q$, $g$ represents the determinant of the metric tensor $g_{\mu \nu}$, and $\mathcal{L}_{m}$ is the Lagrangian density for matter. The trace of the non-metricity tensor $%
Q_{\gamma \mu \nu }$ can be expressed as
\begin{equation}
Q_{\gamma }={{Q_{\gamma }}^{\mu }}_{\mu }\,,\qquad \widetilde{Q}_{\gamma }={%
Q^{\mu }}_{\gamma \mu }\,.
\end{equation}

It is also useful to introduce the superpotential tensor (the conjugate of non-metricity) defined by
\begin{equation}
4{P^{\gamma }}_{\mu \nu }=-{Q^{\gamma }}_{\mu \nu }+2Q{_{(\mu \;\;\nu
)}^{\;\;\;\gamma }}+Q^{\gamma }g_{\mu \nu }-\widetilde{Q}^{\gamma }g_{\mu
\nu }-\delta _{\;(\mu }^{\gamma }Q_{\nu )}\,,
\end{equation}%
where the trace of the non-metricity tensor can be obtained as 
\begin{equation}
Q=-Q_{\gamma \mu \nu }P^{\gamma \mu \nu }\,.
\end{equation}

The field equations of symmetric teleparallel gravity are derived by varying the action $S$ with respect to the metric tensor $g_{\mu \nu}$, resulting in the following equations:
\begin{widetext}
\begin{equation}
\frac{2}{\sqrt{-g}}\nabla _{\gamma }(\sqrt{-g}f_{Q}P^{\gamma }{}_{\mu \nu })+%
\frac{1}{2}fg_{\mu \nu }+f_{Q}(P_{\nu \rho \sigma }Q_{\mu }{}^{\rho \sigma
}-2P_{\rho \sigma \mu }Q^{\rho \sigma }{}_{\nu })=T_{\mu \nu }\;,  \label{F}
\end{equation}%
\end{widetext}
where the energy-momentum tensor is given by 
\begin{equation}
T_{\mu \nu }=-\frac{2}{\sqrt{-g}}\frac{\delta (\sqrt{-g}\mathcal{L}_{m})}{%
\delta g^{\mu \nu }}\;.
\end{equation}
Here $f_{Q}={df}/{dQ}$ and $\nabla _{\mu }$ represents the covariant derivative operator. By varying the action with respect to the connection, we obtain the following equation,
\begin{equation}
\nabla ^{\mu }\nabla ^{\nu }\left( \sqrt{-g}\,f_{Q}\,P^{\gamma }\;_{\mu \nu
}\right) =0.
\end{equation}

The cosmological principle states that our Universe is homogeneous and isotropic on large scales. The mathematical description of a homogeneous and isotropic Universe is given by the flat FLRW metric, which can be expressed as
\begin{equation}
ds^{2}=-dt^{2}+a^{2}(t)\left[ dx^{2}+dy^{2}+dz^{2}\right] ,  \label{FLRW}
\end{equation}%
where $a(t)$ is the scale factor that represents the size of the expanding Universe. The non-metricity scalar corresponding to the FLRW metric is obtained as
\begin{equation}
Q=6H^{2},
\end{equation}%
where $H$ is the Hubble parameter, which represents the rate of expansion of the Universe. To obtain the modified Friedmann equations that govern the Universe when described by the spatially flat FLRW metric, we consider the stress-energy momentum tensor of a perfect fluid, given by
\begin{equation}
T_{\mu \nu }=(p+\rho )u_{\mu }u_{\nu }+pg_{\mu \nu },  \label{PF}
\end{equation}%
where $p$ represents the isotropic pressure, $\rho$ is the energy density, and $u^{\mu}=(1,0,0,0)$ denotes the four-velocity components of the perfect fluid.

In view of Eq. (\ref{PF}) for the spatially flat FLRW metric, the field equations of symmetric teleparallel gravity (\ref{F}) yield the following modified Friedmann equations,
\begin{equation}
3H^{2}=\frac{1}{2f_{Q}}\left( \rho +\frac{f}{2}\right) ,  \label{F1}
\end{equation}%
\begin{equation}
\dot{H}+3H^{2}+\frac{\dot{f}_{Q}}{f_{Q}}H=\frac{1}{2f_{Q}}\left( -p+\frac{f}{2%
}\right) ,  \label{F2}
\end{equation}%
where the dot $(\dot{})$ denotes the derivative with respect to cosmic time $t$. If we choose the function $f(Q)$ to be $f(Q)=Q$, we obtain the standard Friedmann equations \cite{Jimenez2}. This result is expected because, as mentioned earlier, this particular choice of $f(Q)$ corresponds to the theory's limit equivalent to GR. When we instead use $f(Q)=Q+F(Q)$, the field equations (\ref{F1}) and (\ref{F2}) can be expressed as
\begin{equation}
3H^{2}=\rho+\frac{F}{2}-QF_{Q}\,,  \label{F11}
\end{equation}%
\begin{equation}
\left( 2QF_{QQ}+F_{Q}+1\right) \dot{H}+\frac{1}{4}\left( Q+2QF_{Q}-F\right)
=-2p\,,  \label{F22}
\end{equation}
where $F_{Q}=\frac{dF}{dQ}$ and $F_{QQ}=\frac{d^2{F}}{dQ^2}$.

In Eq. (\ref{F11}), we can express the energy density ($\rho$) as the sum of two components, namely, $\rho=\rho_m+\rho_r$, where $\rho_m$ and $\rho_r$ represent the energy densities associated with dark matter and radiation, respectively. Likewise, we can decompose the pressure ($p$) as $p=p_r+p_m$. The conservation equation for standard matter follows as
\begin{equation} \label{eqEoS}
    \frac{d\rho}{dt} + 3 H (1+\omega) \rho=0.
\end{equation}

The Equation of State parameter (EoS) denoted as $\omega$ assumes distinct values depending on the specific matter sources, such as baryonic matter and radiation. In the context of isotropic and homogeneous spatially flat FLRW cosmologies that include radiation, non-relativistic matter, and an exotic fluid characterized by an EoS $p_{de}=\omega_{de} \rho_{de}$, the Friedmann equations \eqref{F11} and \eqref{F22} take on the following form
\begin{equation}\label{eq:8}
    3H^2=\rho_r +\rho_m+\rho_{de},
\end{equation}
\begin{equation}\label{eq:8a}
    2\dot{H}+ 3H^2= -p_r-p_m-p_{de}.
\end{equation}

In this context, $\rho_r$, $\rho_m$, $p_m$, and $p_r$ represent the energy densities of the radiation and matter components, with $p_m\;, p_r$ indicating the pressure associated with matter and radiation. In addition, we have $\rho_{de}$ and $p_{de}$, which represent the density and pressure contributions of DE arising from the geometry, as described by
\begin{equation}
\rho _{de}=\frac{F}{2}-QF_{Q}\,,  \label{F111}
\end{equation}%
\begin{equation}
p_{de}=2\dot{H}(2QF_{QQ}+F_{Q})-\rho _{de}\,.  \label{F222}
\end{equation}

Furthermore, the EoS parameter due to the DE component is 
\begin{equation}
\omega _{de}=\frac{p_{de}}{\rho _{de}\,}=-1+\frac{4\dot{H}(2QF_{QQ}+F_{Q})}{%
F-2QF_{Q}}\,.
\end{equation}

In the following discussion, we make the assumption that the matter pressure, whether it is associated with baryonic matter or dark matter, can be safely disregarded. When there are no interactions between these three distinct fluid components (radiation, non-relativistic matter, and DE), the energy densities obey the following set of differential equations:
\begin{eqnarray}
\dot{\rho}_r+4H\rho_r&=&0,\label{eq:e1}\\
\dot{\rho}_m+3H \rho_m&=&0,\label{eq:e2}\\
\dot{\rho}_{de}+3H(1+\omega_{de})\rho_{de}&=&0 \label{eq:e3}.
\end{eqnarray}

Using Eqs. (\ref{eq:e1}) and (\ref{eq:e2}), it is straightforward to derive the evolution behaviors of pressureless matter and radiation. Specifically, we find that $\rho_m=\rho_{m0} (1+z)^3$, and $\rho_r=\rho_{r0} (1+z)^4$, where $z=\frac{1}{a(t)}-1$ represents the cosmological redshift, and the subscript "0" signifies the value of the respective quantity at the present day or current time.

\section{Cosmological model}

\label{sec3}

For our analysis, we consider a specific functional form of symmetric teleparallel gravity, characterized by the following expression,

\begin{equation}
F\left( Q\right) =\alpha Q^{n},  \label{fQ}
\end{equation}%
where $\alpha \neq 0$ and $n$ are free parameters of the model \footnote{For dimensional consistency, $\alpha$ has the dimensions of $H^{2(1-n)}$.}. Solanki et al. \cite{Solanki} investigated the linear model i.e. $n=1$ in the presence of a viscous fluid. Furthermore, the authors of Refs. \cite{Koussour1,k1} explored this form of gravity within the framework of an anisotropic Universe. When $n=2$, the quadratic form of $f(Q)$ gravity is obtained, and it has been extensively discussed by Koussour et al. \cite{Koussour2} using a hybrid expansion law.

For this general form of $F(Q)$, the modified Friedmann equations, Eqs. (\ref{F111}) and (\ref{F222}), can be expressed as 
\begin{equation}
\rho_{de} =\alpha6^n(\frac{1}{2}- n) H^{2n},  \label{rho}
\end{equation}
\begin{equation}
p_{de}=-\alpha  6^{n-1} (\frac{1}{2}- n) H^{2(n-1)} \left(3 H^2+2n \dot{H}\right).  \label{p}
\end{equation}

Using Eqs. (\ref{rho}) and (\ref{p}), we can express the DE EoS parameter as follows:
\begin{equation}
\omega_{de} =-1-\frac{2n}{3}\left( \frac{\overset{.}{H}}{H^{2}}\right) .
\label{omega}
\end{equation}

The time derivative of the Hubble parameter can be expressed in terms of the cosmological redshift as 
\begin{equation}
\overset{.}{H}=\frac{dH}{dt}=-\left( 1+z\right) H\left( z\right) \frac{dH}{dz}.  \label{dH}
\end{equation}

It is evident that Eqs. (\ref{rho})-(\ref{p}) form a system of two equations with three unknowns: $H$, $\rho_{de}$, and $p_{de}$. Therefore, to solve Eq. (\ref{omega}) for $H(z)$, an additional equation is required. In the literature, an equation for the Hubble parameter is typically employed. However, in this study, we adopt the opposite approach by imposing a constraint on the EoS parameter. This approach is known as the model-independent approach. To investigate DE cosmological models, it is customary to employ a parametrization for relevant variables such as the Hubble or EoS parameters. This parametrization enables us to obtain the necessary equation to solve the field equations. In this study, we consider the Barboza-Alcaniz (BA) parametrization of the DE EoS parameter, which is given by \cite{BA} as 
\begin{equation}
\omega_{de}(z)=\omega_{0}+ \omega_{1}\frac{z(1+z)}{(1+z^2)},
\label{EoS}
\end{equation}%
where $\omega_{0}$ represents the EoS value at the present time, while $\omega _{1}$ quantifies the time dependence of the DE EoS. The BA parametrization exhibits a linear behavior in $z$ at low redshifts, similar to other parameterizations discussed in the literature \cite{BA}. One advantage of this parametrization is its bounded nature, ensuring it remains well-behaved throughout the entire history of the Universe. Furthermore, it demonstrates behavior similar to quintessence and phantom DE models at small redshifts, making it a viable choice for studying the EoS. This parametrization has been extensively discussed in previous studies, including Refs. \cite{Nunes,Wei}. By using Eq. (\ref{EoS}), we can analyze the behavior of the EoS parameter at different redshift values $z$ as follows: 
\begin{itemize}
\item $\omega_{de} =\omega _{0}$, as $z=0$,

\item $\omega_{de} =\omega _{0}+\omega _{1}$, for $z\rightarrow \infty $,

\item $\omega_{de} =\omega _{0}$, for $z\rightarrow -1$.
\end{itemize}

Using Eqs. (\ref{omega}), (\ref{dH}) and (\ref{EoS}), we obtain the Hubble parameter in terms of the cosmological redshift as 
\begin{equation}
H^2\left( z\right) =H_{0}(1+z)^{\frac{3(\omega _{0}+1)}{n}}\left( 1+z^{2}\right) ^{\frac{3\omega_{1}}{2n}},  \label{Hz}
\end{equation}%
where $H_{0}$ represents the present value of the Hubble parameter at $z=0$. It is important to note that the form of the Hubble parameter derived in our study is in agreement with several works in the literature \cite{Hu1,Hu2}.

By substituting \eqref{Hz} into \eqref{rho}, we derive the expression for the DE density $\rho_{de}$ as
\begin{equation}
\rho_{de}(z)=\alpha6^n(\frac{1}{2}- n) H_{0}^{2n} (z+1)^{3 (\omega_0+1)} \left(z^2+1\right)^{\frac{3 \omega_1}{2}}.
\end{equation}

Thus, the Friedmann equation (\ref{eq:8}) can be expressed as follows:
\begin{widetext}
\begin{equation}
\frac{H^2(z)}{ H_0^2}=\Omega_{r0}(1+z)^4+\Omega_{m0}(1+z)^3+ \Omega_{de0}(z+1)^{3 (\omega_0+1)} \left(z^2+1\right)^{\frac{3 \omega_1}{2}}\;,
\end{equation}
\end{widetext}
where we have defined, for this particular model of $f(Q)$,
\begin{equation}
    \Omega_{de0}\equiv\frac{\alpha6^n}{3}(1/2-n)H_0^{2(n-1)}\;.
\end{equation}


Furthermore, we have $\Omega _{r0}$ and $\Omega _{m0}$ representing the present-day values of the radiation and matter density parameters, defined as $\Omega _{r0}=\frac{\rho_{r0}}{3H{0}^{2}}$ and $\Omega _{m0}=\frac{\rho_{m0}}{3H{0}^{2}}$, respectively.

\section{Observational constraints}
\label{sec4} 

In this section, we proceed to constrain our model parameters by comparing them with the Hubble ($Hz$) and $Pantheon$ datasets. The best values of the parameters, along with their uncertainties, are determined through the utilization of the MCMC method \cite{Mackey} and by minimizing the chi-square function $\chi^2$. To assess the goodness of fit, we calculate the total $\chi^2$ by combining the contributions from both the $Hz$ and $Pantheon$ samples. The expression for the total $\chi^2$ is given by,
\begin{equation}
\chi^2_{tot}=\chi^2_{Hz}+\chi^2_{\text{Pantheon}},
\end{equation}
where $\chi^2_{Hz}$ represents the chi-square value associated with the $Hz$ measurements, and $\chi^2_{\text{Pantheon}}$ corresponds to the chi-square value of the $Pantheon$ dataset.

\subsection{$Hz$ dataset}

To begin, we use a standard compilation of 31 $Hz$ data measurements acquired via the differential age method \cite{Yu/2018,Moresco/2015}. This method allows for the estimation of the universe's expansion rate at a given redshift $z$. Specifically, $H(z)$ can be calculated as $H(z)=-\frac{dz/dt}{(1+z)}$. The $
\chi_{H}^2$ function is defined as 
\begin{equation}
    \chi^{2}_{Hz}= \sum_{i=1}^{31} \frac{\left[H(z_{i}, \mathcal{P})-H_{obs}(z_{i})\right]^{2} }{\sigma(z_{i})^{2}}\;,
\end{equation}
where $H(z_{i}, \mathcal{P})$ represents the theoretical value of the model at redshifts $z_{i}$, and $\mathcal{P}$ denotes the parameter space, namely $H_0$, $\Omega_{m0}$, $\alpha$, $\omega_{0}$, $\omega_{1}$, $n$. On the other hand, $H_{obs}(z_{i})$ and $\sigma(z_{i})^2$ correspond to the observed value and the error, respectively.

\subsection{$Pantheon$ dataset}

Secondly, we use a dataset comprising 1048 data points from the $Pantheon$ compilation, which consists of SN Ia observations. These data points span the redshift range $0.01\leqslant z\leqslant2.3$ \cite{Scolnic}. The $Pantheon$ sample combines data from different supernova surveys such as SDSS, SNLS, various low-z samples, and high-z samples from HST. The corresponding chi-square $\chi^2_{Pantheon}$ for the $Pantheon$ dataset is defined as
\begin{equation}
\chi^{2}_{Pantheon}= \sum_{i,j=1}^{1048} \Delta \mu_{i} (C^{-1}_{Pantheon})_{ij} \Delta \mu_{j}, 
\end{equation}
where $\mu= \mu_{obs}(z_{i})-\mu_{th}(\mathcal{P},z_{i})$. Here, $\mu_{obs}(z_{i})$ is the observational distance
modulus, $\mu_{th}(\mathcal{P},z_{i})$ is the theoretical value defined
as 
\begin{equation}
\mu_{th}(\mathcal{P},z_{i}) = 5log_{10}\left(\frac{d_{L}(z)}{1 Mpc}\right)+25,
\label{mu}
\end{equation}%
and $C^{-1}_{Pantheon}$ is the inverse covariance matrix. In addition, the luminosity distance $d_{L}(z)$ in Eq. (\ref{mu}) is defined as 
\begin{equation}
d_{L}(z)= c(1+z) \int_{0}^{z} \frac{dz'}{H(z',\mathcal{P})},
\end{equation}%
where $c$ is the speed of light.

The $1-\sigma$ and $2-\sigma$ contours on the model parameters $H_0$, $\Omega_{m0}$, $\alpha$, $\omega_{0}$, $\omega_{1}$, $n$ are presented in Fig. \ref{fig_triangle}, and the corresponding numerical results are summarized in Tab. \ref{tab}. Figs. \ref{fig_H} and \ref{fig_Mu} show a comparison between our cosmological model and the standard $\Lambda$CDM model. For this comparison, we adopted the values $\Omega_{m0} = 0.315$ and $H_0 = 67.4\pm0.5$ km/s/Mpc, which were obtained from recent measurements by the Planck satellite \cite{Planck2018}. The figures display the data points of the Hubble parameter (31 data points) and the Pantheon compilation (1048 data points) along with their corresponding error bars. We can observe that our model provides a good fit for the data. By minimizing the $\chi^2$ function with respect to the mode parameters $(H_0,\Omega_{m0},\alpha,\omega_{0},\omega_{1},n)$, we obtain the best-fit values $\Omega_{m0}=0.23_{-0.22}^{+0.20}$, $\alpha=-0.59_{-0.40}^{+0.39}$, $\omega_{0}=-1.01_{-0.48}^{+0.39}$, $\omega_{1}=0.48_{-0.23}^{+0.23}$, $n=1.028_{-0.071}^{+0.082}$ for the $Hz+Pantheon$ dataset (see Tab. \ref{tab}). These parameter values result in a best-fit value for the present Hubble parameter of $H_0 = 68.0_{-8}^{+10}$. Remarkably, our model reduces the Hubble tension compared to the value obtained by the SH0ES project, $H_0 = 73.2\pm1.3$ $km/s/Mpc$ at $68\%$ confidence level \cite{Riess2021}. In our model, we have selected the $f(Q)$ function as $f(Q) = Q + F(Q)$, where $F(Q) = \alpha Q^n$. It is essential to note that to achieve what is referred to as the Symmetric Teleparallel Equivalent to GR, we set $F(Q) = 0$, which implies $\alpha = 0$. However, it is not equivalent to the $\Lambda$CDM model, as the cosmological constant is not present in this case. To obtain the $\Lambda$CDM model, we set $n = 0$, and $\alpha = 2\Lambda$ \cite{Khyllep}. The deviations from $n=0$ introduce modifications to the model that go beyond the GR framework and give rise to the DE component within the $f(Q)$ model and, consequently, the differences observed in our analysis. Specifically, it leads to variations in the cosmic expansion scenario and provides a framework for exploring alternative cosmological dynamics. Also, it is very important to note that we have omitted the radiation density parameter, $\Omega_{r0}$, due to its negligible contribution in comparison to other dominant components, and its omission does not significantly impact the results of the MCMC analysis or the conclusions of this study.

\begin{widetext}

\begin{figure}[h]
\centerline{\includegraphics[scale=0.50]{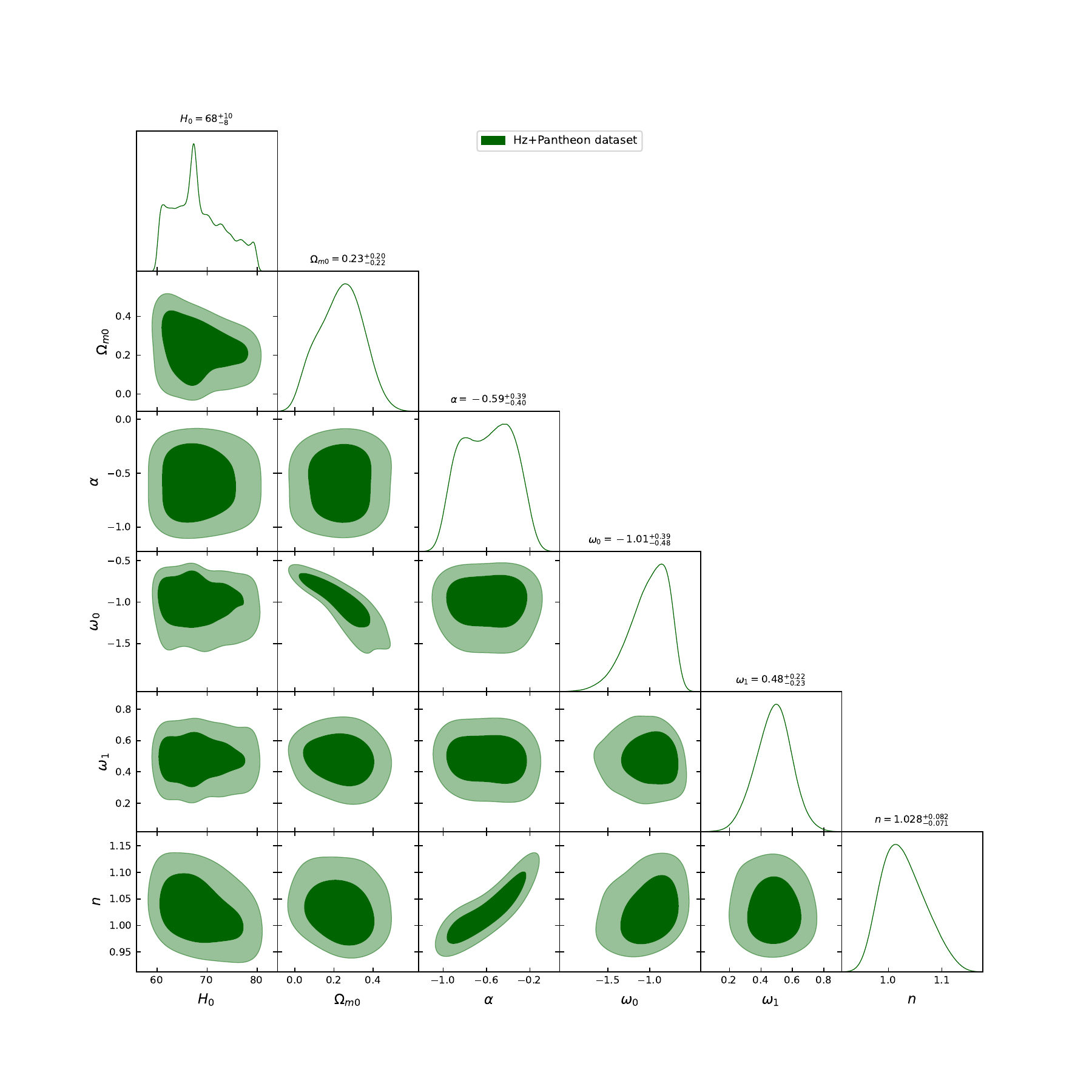}}
\caption{The $1-\sigma $ and $2-\sigma $\ contours for the model parameters $H_0$, $\Omega_{m0}$, $\alpha$, $\omega_{0}$, $\omega_{1}$, $n$ using $Hz+Pantheon$ dataset.}
\label{fig_triangle}
\end{figure}

\begin{figure}[h]
\centerline{\includegraphics[scale=0.50]{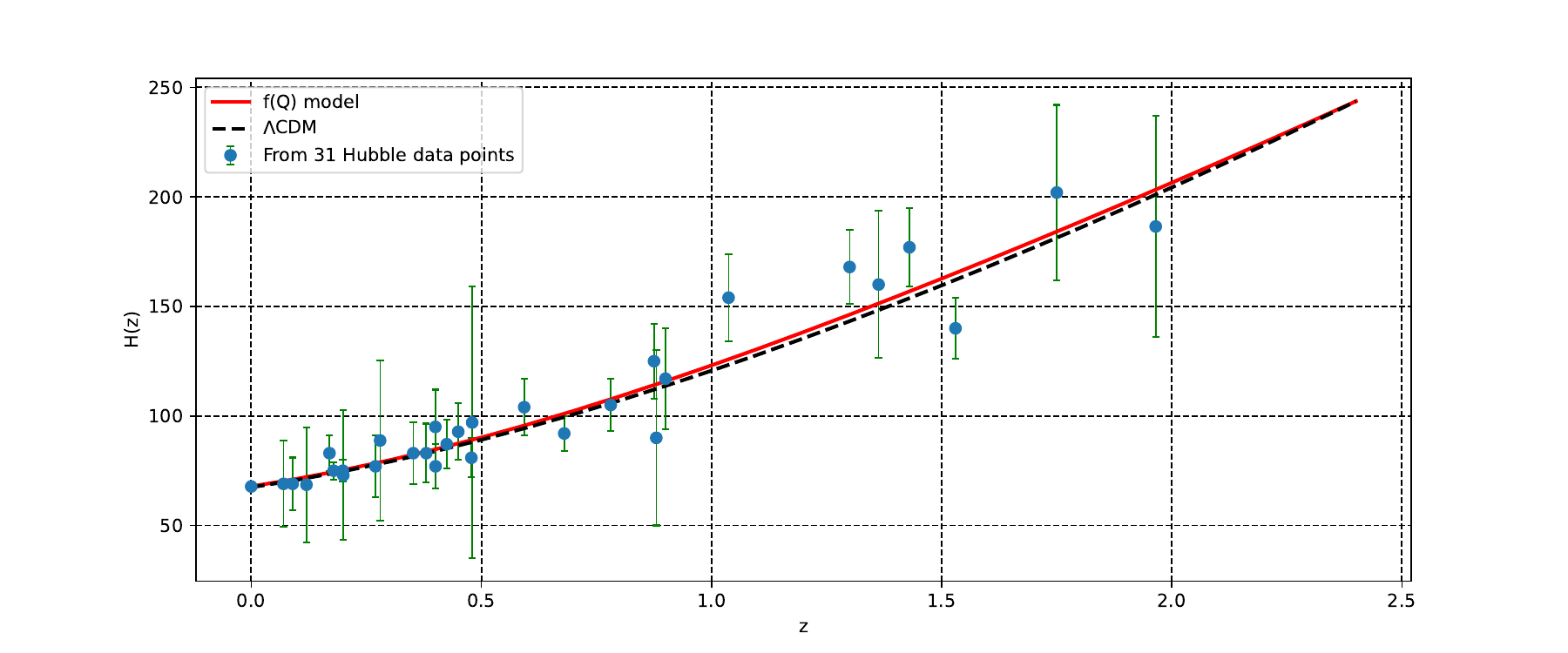}}
\caption{The evolution of the Hubble parameter $H(z)$ with redshift $z$:
Comparison between our $f(Q)$ cosmological model (red dashed line) and the $\Lambda$CDM model (black dashed line) alongside observed $H(z)$ data points (green dots) with error bars.}
\label{fig_H}
\end{figure}

\begin{figure}[h]
\centerline{\includegraphics[scale=0.50]{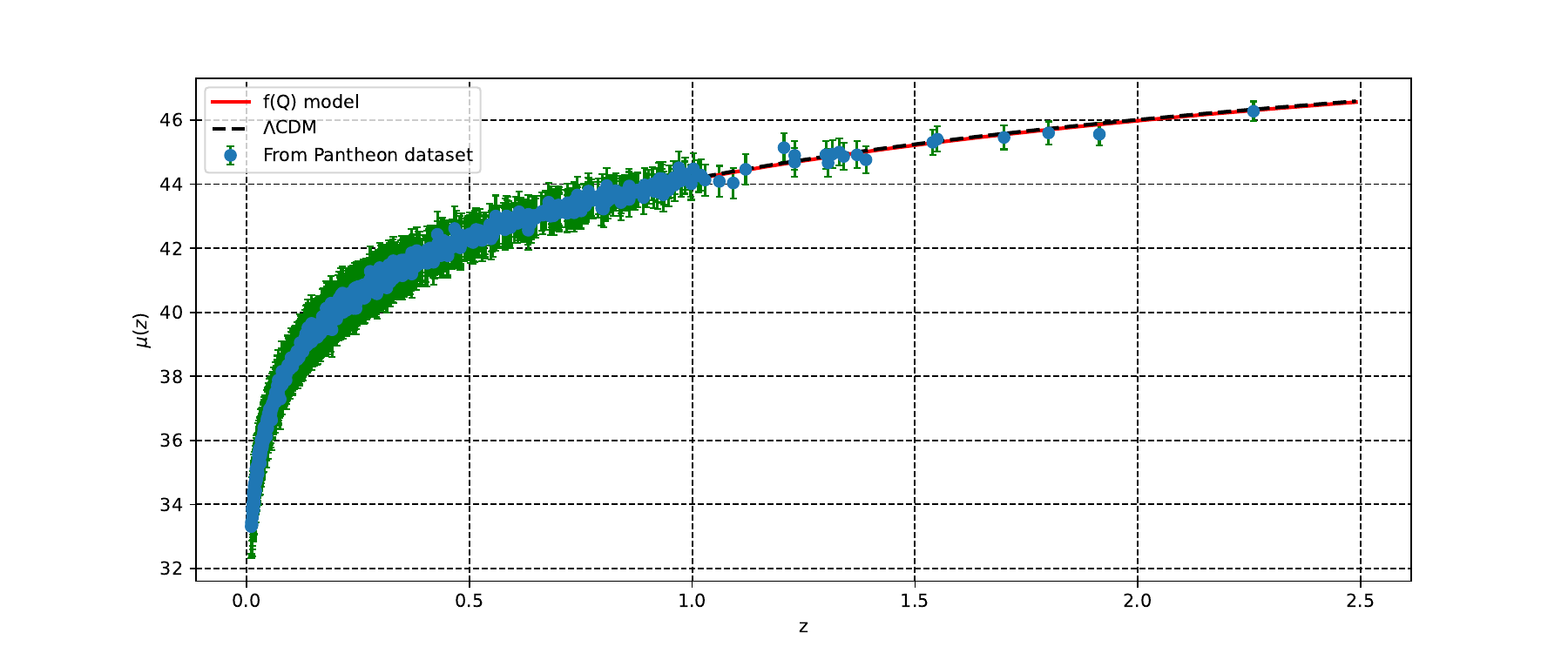}}
\caption{The evolution of the distance modulus $\mu(z)$ with redshift $z$:
Comparison between our $f(Q)$ cosmological model (red dashed line) and the $\Lambda$CDM model (black dashed line) alongside observed $Pantheon$ data points (green dots) with error bars.}
\label{fig_Mu}
\end{figure}

\end{widetext}

\begin{table*}[!htbp]
\begin{center}
\begin{tabular}{l c c c c c c}
\hline\hline 
$datasets$              & $H_0$ & $\Omega_{m0}$ & $\alpha$ & $\omega_{0}$ & $\omega_{1}$ & $n$ \\
\hline
$Priors$   & $(60,80)$  & $(0,1)$  & $(-1,1)$ & $(-2,2)$ & $(-2,2)$ &$(-10,10)$\\

$Hz+Pantheon$   & $68.0_{-8}^{+10}$ & $0.23_{-0.22}^{+0.20}$  & $-0.59_{-0.40}^{+0.39}$ & $-1.01_{-0.48}^{+0.39}$ & $0.48_{-0.23}^{+0.23}$ & $1.028_{-0.071}^{+0.082}$\\

\hline\hline
\end{tabular}
\caption{The best-fit values of the model parameters using $Hz+Pantheon$ dataset. Also shown are the present values of the cosmological parameters.}
\label{tab}
\end{center}
\end{table*}

\subsection{The deceleration parameter}

The deceleration parameter, a fundamental concept in cosmology, is pivotal for understanding the dynamics of the universe's expansion. It is defined mathematically in relation to the Hubble parameter as follows:
\begin{equation*}
q=-1-\frac{\overset{.}{H}}{H^{2}}.
\end{equation*}

This parameter plays a central role in cosmological models, as it characterizes whether the Universe's expansion is accelerating or decelerating. When $q$ is negative ($q<0$), it signifies an accelerating expansion, as observed in the case of DE-dominated Universes. Conversely, when $q$ is positive ($q>0$), it indicates a decelerating expansion, as typically seen in matter-dominated Universes. Understanding $q$ is crucial to gain insight into the past, present, and future evolution of our Universe. The observational data employed in this study provide evidence that our present Universe has entered an accelerating phase, with the deceleration parameter lying within the range of $-1\leq q< 0$. In our analysis, we can express the deceleration parameter in terms of the cosmological parameters employed as
\begin{equation}
q(z)=\Omega_r(z)+\frac{1}{2}\Omega_m(z)+\frac{1+3\omega_{de}}{2} \Omega_{de}(z).  \label{DP}
\end{equation}

\begin{figure}[h]
\centerline{\includegraphics[scale=0.75]{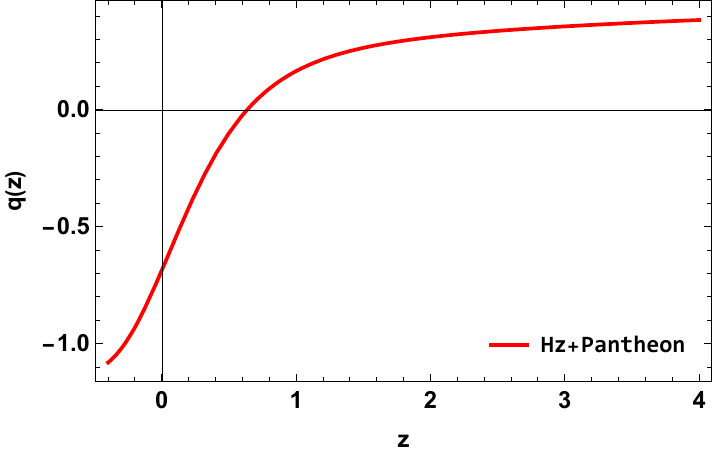}}
\caption{The evolution of deceleration parameter with redshift $z$ using constraints from the $Hz+Pantheon$ dataset.}
\label{fig_q}
\end{figure}

From the analysis presented in Fig. \ref{fig_q}, it is clear that the deceleration parameter captures the two distinct phases of the Universe: the deceleration phase and the subsequent acceleration phase, which have been observed in various studies \cite{Acc1, Acc2}.  In our model, the transition between these phases occurs at a redshift value of $z_{tr}=0.64^{+0.07}_{-0.07}$ \cite{Jesus,Garza}, determined using the $Hz+Pantheon$ dataset. Moreover, the present value of the deceleration parameter is $q_{0}=-0.69^{+0.59}_{-0.66}$ \cite{Hernandez,Mamon}. This negative value aligns with the observed acceleration phase of the Universe, further supporting the validity of our model.

\subsection{The EoS parameter}

The EoS parameter is a fundamental quantity that provides insights into the properties of proposed DE models. It is defined as the ratio of the isotropic pressure $p$ to the energy density $\rho$ of the Universe, given by $\omega = \frac{p}{\rho}$. In order to explain the observed cosmic acceleration, it is necessary for the EoS parameter to satisfy $\omega < -\frac{1}{3}$. This condition ensures that the dominant component of the Universe's energy density possesses negative pressure, which drives the accelerated expansion. The simplest and most widely studied candidate for DE is the cosmological constant $\Lambda$ in the framework of GR. It has a constant EoS parameter given by $\omega_{\Lambda} = -1$. This value indicates that the cosmological constant behaves like a fluid with negative pressure, causing a repulsive gravitational effect that leads to cosmic acceleration. However, there are alternative dynamical models of DE, such as quintessence, where the EoS parameter lies in the range $-1 < \omega_{de} < -\frac{1}{3}$. These models introduce a dynamical scalar field that evolves with time and can mimic the behavior of DE. Another intriguing possibility is phantom energy, characterized by an EoS parameter $\omega_{de} < -1$. In this case, the energy density increases with time, leading to a super-accelerated expansion and potential future cosmic singularities.

\begin{figure}[h]
\centerline{\includegraphics[scale=0.75]{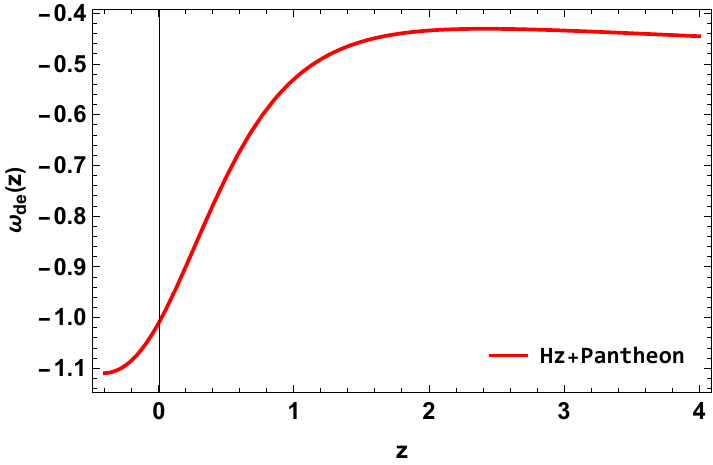}}
\caption{The evolution of effective EoS parameter with redshift $z$ using constraints from the $Hz+Pantheon$ dataset.}
\label{fig_omega}
\end{figure}

The behavior of the effective EoS parameter is depicted in Fig. \ref{fig_omega}, where we present the results obtained from analyzing the $Hz+Pantheon$ datasets. It is evident from the figure that the DE EoS parameter of our analysis exhibits phantom-like behavior, characterized by $\omega_{de} < -1$. This indicates that the dominant component responsible for the accelerated expansion of the Universe behaves in a manner similar to phantom models of DE. Furthermore, we find that the present value of the DE EoS parameter corresponding to the $Hz+Pantheon$ dataset is $\omega_0 =-1.01_{-0.48}^{+0.39}$ (see Tab. \ref{tab}) \cite{Feng, Novosyadlyj/2012, Suresh/2014}. This value suggests that the current cosmic acceleration is well described by our model. The negative value of $\omega_0$ indicates that the Universe is currently experiencing an accelerated expansion, consistent with the observational data. 

\begin{figure}[h]
\centerline{\includegraphics[scale=0.75]{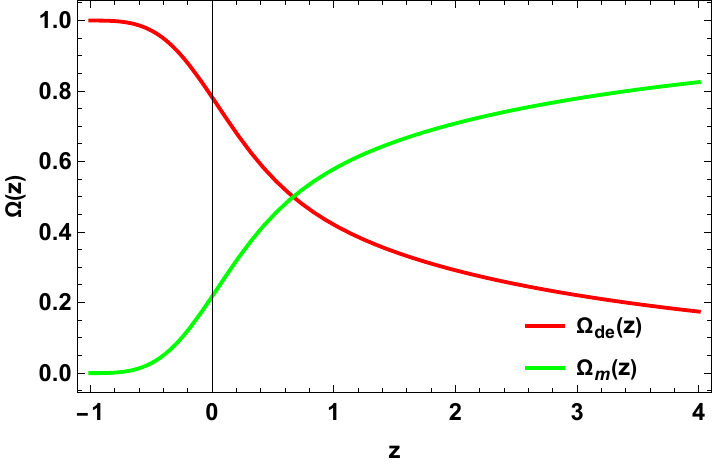}}
\caption{The evolution of density parameter with redshift $z$ using constraints from the $Hz+Pantheon$ dataset.}
\label{fig_rho}
\end{figure}

The energy density sources in our universe exhibit dynamic evolution over time, and they play a pivotal role in defining cosmic history, its present state, and its future prospects. In Fig. \ref{fig_rho}, we have provided insightful visualizations of the evolving DE density and matter density. From this graphical representation, it becomes evident that in the early epochs, the matter density was the dominant force shaping the Universe, while in the present phase, the DE density holds sway, even contributing to the current acceleration of cosmic expansion. In the context of the $Hz+Pantheon$ dataset, we have determined the present-day value of the matter density to be approximately $0.23$, with a $1-\sigma$ error range of $+0.20$ and $-0.22$. Detailed constraint values for the matter density at the 68\% and 95\% confidence levels are also tabulated in Tab. \ref{tab}. Furthermore, a noteworthy observation is that throughout the entire course of their evolution, the sum of the matter density and DE density remains remarkably close to unity ($\Omega_m+\Omega_{de}\simeq 1$), reflecting a critical balance in the cosmic energy budget. These dynamic profiles of the two energy components strongly suggest that DE is poised to continue its dominance in our Universe's foreseeable future, further contributing to its intriguing and complex cosmic story.

\section{Statefinder analysis}
\label{sec5}

The deceleration and EoS parameters are important in characterizing the expansion and nature of the Universe. However, a challenge arises because many proposed DE models in the literature share the same current values for these parameters. Consequently, these parameters are not sufficient to effectively distinguish between the different models under study. To address this issue, Sahni et al. \cite{Sahni, Alam} introduced a new pair of dimensionless cosmological parameters called statefinder parameters $(r, s)$, which offer a more discriminating diagnostic for DE models. The statefinder parameters are defined as
\begin{equation}
r=\frac{\overset{...}{a}}{aH^{3}},
\end{equation}
\begin{equation}
s=\frac{\left( r-1\right) }{3\left( q-\frac{1}{2}\right) }.
\end{equation}

The parameter $r$ can be expressed in terms of the deceleration parameter as
\begin{equation}
r=2q^{2}+q-\frac{\overset{.}{q}}{H}.
\end{equation}

The trajectories in the $r-s$ plane are important for classifying different cosmological regions, and various DE models can be characterized using this diagnostic pair as:
\begin{itemize}
\item $\Lambda$CDM model corresponds to ($r=1,s=0$),

\item The holographic DE model corresponds to ($r=1,s=\frac{2}{3}$),

\item Chaplygin gas model corresponds to ($r>1,s<0$),

\item Quintessence model corresponds to ($r<1,s>0$),
\end{itemize}

\begin{figure}[h]
\centerline{\includegraphics[scale=0.75]{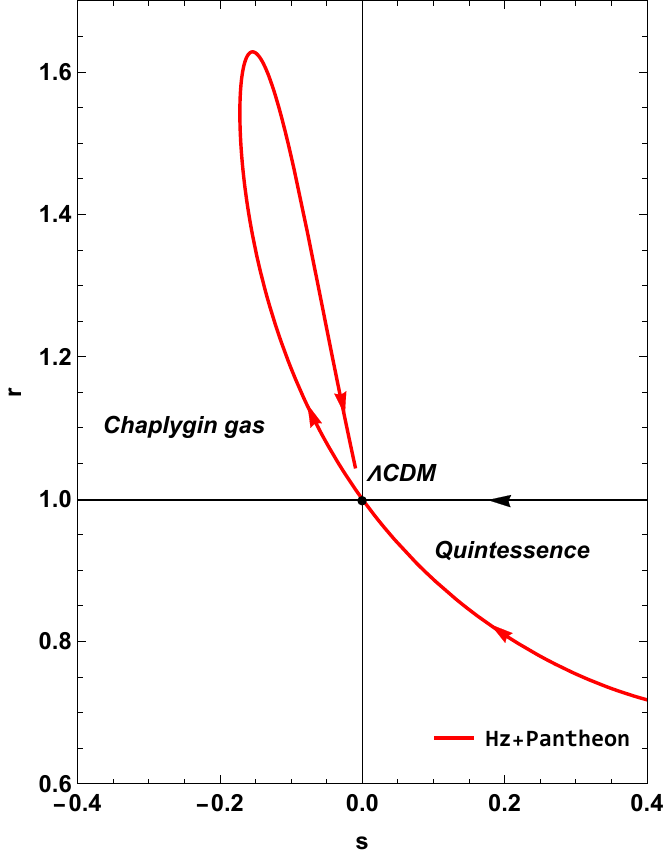}}
\caption{The evolution of the $r-s$ plane using constraints from the $Hz+Pantheon$ dataset.}
\label{fig_rs}
\end{figure}

Fig. \ref{fig_rs} represents the $r-s$ plane, where the parameters are constrained by the $Hz+Pantheon$ dataset. The plot provides valuable insights into the behavior of these parameters over cosmic time. Notably, it becomes evident that, in the early universe, the parameter values satisfy conditions $r < 1$ and $s > 0$. These conditions suggest that the DE candidate in our model exhibits quintessence-like behavior during these early epochs. However, as we transition to the present epoch (at $z=0$), the model manifests different characteristics. Furthermore, in the late-time cosmic regime, as $z$ approaches $-1$, the model adopts properties akin to the $\Lambda$CDM model. This intriguing result corroborates our earlier findings concerning the EoS parameter, reinforcing the notion that the behavior of DE in our model undergoes distinct phases, resembling quintessence at early times and converging towards $\Lambda$CDM-like behavior in the late Universe.

\section{$Om(z)$ diagnostic}
\label{sec6}

In this section, we introduce another valuable tool for investigating the dynamic nature of cosmological models pertaining to DE, known as the Om diagnostic \cite{Sahni1}. This diagnostic offers a simpler approach compared to the statefinder diagnostic discussed earlier, as it relies solely on the Hubble parameter $H$. In a spatially flat Universe, the $Om(z)$ diagnostic is defined as 
\begin{equation}
Om\left( z\right) =\frac{E^{2}\left( z\right) -1}{\left( 1+z\right) ^{3}-1}\;,
\end{equation}%
where $E\left( z\right) =\frac{H\left( z\right) }{H_{0}}$. The behavior of $Om(z)$ provides valuable information on the nature of DE in the cosmological model. A negative slope of $Om(z)$ indicates quintessence behavior, where the energy density of DE decreases with time. On the other hand, a positive slope represents a phantom behavior, where the energy density increases with time. A constant value of $Om(z)$ corresponds to the standard $\Lambda$CDM model. 

\begin{figure}[h]
\centerline{\includegraphics[scale=0.75]{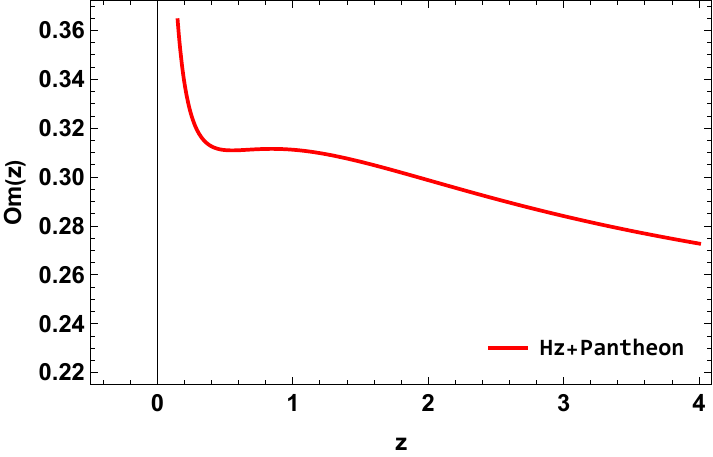}}
\caption{The evolution of $Om(z)$ diagnostic with redshift $z$ using constraints from the $Hz+Pantheon$ dataset.}
\label{fig_om}
\end{figure}

The plot presented in Fig. \ref{fig_om} offers compelling insights into the behavior of the $Om(z)$ diagnostic as a function of redshift $z$. Notably, it becomes apparent that for $z<0$,
$Om(z)$ displays a negative slope. This intriguing observation suggests that in the early Universe, our cosmological model indeed showcases quintessence-like characteristics for DE. In quintessence, the EoS of DE $\omega_{de}$ lies between $-1$ (indicating a cosmological constant) and $-1/3$ (representing matter-like behavior), which is consistent with the negative slope of $Om(z)$ at these redshifts. However, as we extend our view to late cosmic times, a distinct transformation occurs. In this late-time regime, when $z$ approaches values close to $-1$, our model adopts properties akin to phantom-like DE. Phantom DE corresponds to $\omega_{de}<-1$, and it is associated with an expanding universe that accelerates at an increasing rate. The shift towards phantom-like behavior in the late Universe is a fascinating feature of our cosmological model, highlighting its capacity to encompass diverse phases of DE evolution, from quintessence-like to phantom-like, as the cosmic epoch unfolds.

\section{Conclusions}
\label{sec7}

In this paper, we investigate the late-time acceleration of the Universe within the framework of $f(Q)$ gravity. In our model, we adopt a parametrization of the EoS to describe the behavior of DE. By performing an MCMC analysis using observational data from $H(z)$ measurements and the $Pantheon$ dataset, we derive the best-fit parameters for our model (see Tab. \ref{tab}). Our MCMC analysis yields results that are consistent with the current understanding of the accelerated expansion of the Universe. Specifically, we have found that the Universe experiences a transition from a deceleration phase to an acceleration phase at a redshift of $z_{tr}=0.64^{+0.07}_{-0.07}$ for $Hz+Pantheon$ dataset \cite{Jesus,Garza}. This transition is supported by a negative value of the deceleration parameter i.e. $q_{0}=-0.69^{+0.59}_{-0.66}$ \cite{Hernandez,Mamon}, indicating the onset of cosmic acceleration.

Furthermore, the EoS parameter exhibits a phantom-like behavior for DE ($\omega_{de} < -1$). The presence of phantom DE in a cosmological model leads to some intriguing consequences \cite{Caldwell/2003}. Specifically, it implies that as the universe expands, the rate of cosmic acceleration increases over time, ultimately leading to a "Big Rip" scenario \cite{McInnes}. In the Big Rip, the universe's expansion becomes so rapid that it tears apart not only galaxies, stars, and planets but even atoms themselves, resulting in a catastrophic end to the cosmos. Also, our analysis of the $Hz+Pantheon$ dataset reveals that the present value of the EoS parameter is $\omega_0 =-1.01_{-0.48}^{+0.39}$ \cite{Feng, Novosyadlyj/2012, Suresh/2014}. We also utilized two diagnostic tools, the statefinder and the $Om(z)$ diagnostic, to further investigate the properties of DE in our model. Ultimately, our comprehensive analysis consistently reinforces the notion of a phantom-like behavior for DE as the driving force behind the late-time acceleration of the Universe. This conclusion finds support across multiple cosmological diagnostics, including the deceleration parameter, EoS parameter, statefinder, and the $Om(z)$ diagnostic. Collectively, these diagnostic tools converge to provide a compelling portrait of the evolving nature of DE, unveiling its phantom-like attributes in the cosmic story's late chapters.

\section*{Acknowledgments}
The authors would like to thank the Deanship of Scientific Research at Imam Mohammad Ibn Saud Islamic University (IMSIU) for supporting this work by: (grant number IMSIU-RG23008).

\textbf{Data availability} There are no new data associated with this
article.


\end{document}